\documentclass{article}

\usepackage{amsmath}
\usepackage{amssymb}

\numberwithin{equation}{section}

\begin{document}
\bibliographystyle{unsrt}

\title{A lower bound for the eigenvalues of the Sen--Witten operator 
on closed spacelike hypersurfaces}

\author{ L\'aszl\'o B Szabados \\
Research Institute for Particle and Nuclear Physics \\
H-1525 Budapest 114, P. O. Box 49, Hungary \\
e-mail: lbszab@rmki.kfki.hu }
\maketitle

\begin{abstract}
The eigenvalue problem for the Sen--Witten operator on closed spacelike 
hypersurfaces is investigated. The (square of its) eigenvalues are 
shown to be given exactly by the 3-surface integral appearing in the 
expression of the total energy-momentum of the matter+gravity systems 
in Witten's energy positivity proof. A sharp lower bound for the 
eigenvalues, given in terms of the constraint parts of the spacetime 
Einstein tensor, i.e. the energy and momentum densities of the matter 
fields, is given.

\end{abstract}


\section{Introduction}
\label{sec-1}

A promising approach of constructing observables of the gravitational 
field in general relativity could be based on the spectral analysis of 
the Dirac operators on various submanifolds of the spacetime. For 
example, the eigenvalues of these operators are such gauge invariant 
objects, which are expected to reflect the geometrical properties of 
the submanifold in question, e.g. in the form of some lower bound for 
the eigenvalues in terms of other well known geometrical objects. 
The first who gave such a lower bound was Lichnerowicz \cite{L}: he 
showed, in particular, that on a closed Riemannian spin manifold 
$\Sigma$ with positive scalar curvature $\frac{1}{4}\inf\{ R(p)\vert 
p\in\Sigma\}$ is a lower bound for the square of the eigenvalues. 
However, this bound is not sharp: on a metric 2-sphere with radius 
$r$ the (positive) eigenvalues are $\frac{n}{r}$, $n\in{\mathbb N}$, 
while on metric spheres the bounds were expected to be saturated. In 
fact, in the last two decades such {\em sharp} lower bounds were found 
in terms of the scalar curvature \cite{TFr,Hi86,Hi95,TF00} or the 
volume \cite{Ba92,TF00,FK}. In particular, in dimension $m$ the sharp 
lower bound, given by Friedrich \cite{TFr,TF00}, is $\frac{m}{4(m-1)}
\inf\{ R(p)\vert p\in\Sigma\}$. Similar results exist for hypersurface 
Dirac operators when the lower bounds are given in terms of the 
curvature scalar of the intrinsic geometry and the trace of the 
extrinsic curvature \cite{Zh}. 

To have significance of these results in general relativity we should 
be able to link the bounds to well known concepts of physics, e.g. the 
objects defined in a natural way on a spacelike hypersurface $\Sigma$ 
of a Lorentzian 4-manifold. For example, the curvature scalar $R$ of 
the intrinsic geometry of $\Sigma$, by means of which many of the 
bounds above were formulated, or the square of the trace $\chi$ of 
the extrinsic curvature of $\Sigma$ in the spacetime, appearing in the 
bound given in \cite{Zh}, are not really `4-covariant'. They are only 
terms in the Hamiltonian constraint part of the spacetime Einstein 
tensor. Moreover, the sign of $\chi^2$ in the bound given in \cite{Zh} 
is {\em negative}, which {\em decreases} the lower bound, and hence its 
usefulness is questionable. 

The aim of the present paper is to derive a sharp lower bound for the 
eigenvalues of the Sen--Witten operator (i.e. the Dirac operator built 
from the hypersurface Sen connection) on closed spacelike 
hypersurfaces of the spacetime, which bound has a clear physical 
interpretation. We give a new kind of lower bound, given in terms of 
the constraint parts of the four dimensional Einstein tensor, rather
than the intrinsic scalar curvature or the square of the trace of the 
extrinsic curvature. Its physical significance comes from the fact 
that, through Einstein's equations, this is just the energy and 
momentum density of the matter fields, for which we have a certain 
form of a positivity requirement (dominant energy condition). 
We find that on a closed spacelike hypersurface $\Sigma$ the 
eigenvalues of the Sen--Witten operator is given by the expression of 
the total energy of the matter+gravity systems appearing in Witten's 
positive energy proof. This provides a sharp lower bound for the 
eigenvalues: it is an average of the flux of the energy current of 
the matter fields seen by the null observers. Through the example of 
a $t={\rm const}$ hypersurface of the $k=1$ Friedman--Robertson--Walker 
spacetime we show that this bound is sharp. 

We use the abstract index formalism and the sign conventions of 
\cite{PRI}. In particular, the signature of the spacetime metric is 
$(+,-,-,-)$, the curvature and Ricci tensors and the curvature scalar 
are defined by $R^a{}_{bcd}X^b:=-(\nabla_c\nabla_d-\nabla_d\nabla_c)
X^a$, $R_{bd}:=R^a{}_{bad}$ and $R:=R_{ab}g^{ab}$, respectively. 
Then Einstein's equations take the form $G_{ab}=-\kappa T_{ab}$, 
where $\kappa:=8\pi G$ with Newton's gravitational constant $G$.


\section{Geometrical preliminaries}
\label{sec-2}

\subsection{Metrics on bundles over $\Sigma$}
\label{sub-2.1}

Let $\Sigma$ be a smooth orientable spacelike hypersurface, $t^a$ 
its future pointing unit normal, and define $P^a_b:=\delta^a_b-t^a
t_b$. This is the orthogonal projection to $\Sigma$, by means of 
which the induced (negative definite) 3-metric is defined by $h
_{ab}:=P^c_aP^d_bg_{cd}$. We assume that the spacetime is space 
and time orientable, at least on an open neighbourhood of $\Sigma$, 
in which case $t^a$ can be (and, in what follows, will be) chosen 
to be globally defined. 

Let ${\mathbb V}^a(\Sigma)$ denote the pull back to $\Sigma$ of the 
spacetime tangent bundle, which decomposes in a unique way to the 
$g_{ab}$-orthogonal direct sum of the tangent bundle $T\Sigma$ and 
the normal bundle spanned by $t^a$. $g_{ab}$ is a Lorentzian fibre 
metric, and we call the triple $({\mathbb V}^a(\Sigma),g_{ab},P^a_b)$ 
the Lorentzian vector bundle over $\Sigma$. It is the projection 
$P^a_b$ as a base point preserving bundle endomorphism which tells 
us how the tangent bundle $T\Sigma$ is embedded in ${\mathbb V}^a
(\Sigma)$. Since both $T\Sigma$ and the normal bundle of $\Sigma$ 
in $M$ are globally trivializable, ${\mathbb V}^a(\Sigma)$ is also. 
This implies the existence of a spinor structure too. Let ${\mathbb 
S}^A(\Sigma)$ denote the bundle of 2-component (i.e. Weyl) spinors 
over $\Sigma$, and we denote the complex conjugate bundle by $\bar
{\mathbb S}^{A'}(\Sigma)$. As is usual in general relativity (see 
e.g. \cite{PRI}), we identify the Hermitian subbundle of ${\mathbb 
S}^A(\Sigma)\otimes\bar{\mathbb S}^{A'}(\Sigma)$ with ${\mathbb V}
^a(\Sigma)$. Thus we can convert tensor indices to pairs of spinor 
indices and vice versa freely. 

On the spinor bundle two metrics are defined: The first is the 
natural symplectic metric $\varepsilon_{AB}$, while the other is the 
positive definite Hermitian metric $G_{AB'}:=\sqrt{2}t_{AB'}$. 
(The reason of the factor $\sqrt2$ is that for this definition $G
^{AB'}$, the inverse of $G_{AB'}$ defined by $G^{AB'}G_{BB'}=\delta
^A_B$, is just the contravariant form $\varepsilon^{AC}\varepsilon
^{B'D'}G_{CD'}$ of the Hermitian metric, i.e. the Hermitian and the 
symplectic metrics are compatible.) 
The Hermitian metric defines the ${\mathbb C}$-linear bundle 
isomorphisms $\bar{\mathbb S}^{A'}(\Sigma)\rightarrow{\mathbb S}^A
(\Sigma):\bar\lambda^{A'}\mapsto-G^A{}_{A'}\bar\lambda^{A'}$ and 
$\bar{\mathbb S}_{A'}(\Sigma)\rightarrow{\mathbb S}_A(\Sigma):\bar
\lambda_{A'}\mapsto G_A{}^{A'}\bar\lambda_{A'}$; as well as 

\begin{equation}
\langle\lambda_A,\phi_A\rangle:=\int_\Sigma G^{AA'}\lambda_A\bar\phi
_{A'}{\rm d}\Sigma, \label{eq:2.1}
\end{equation}
which is a global $L_2$ scalar product on the space of the (square 
integrable) spinor fields on $\Sigma$. This defines a norm in the 
standard way: $\Vert\lambda_A\Vert^2:=\langle\lambda_A,\lambda_A
\rangle$. 


\subsection{The Sen connection}
\label{sub-2.2}

The intrinsic Levi-Civita covariant derivative operator, defined on 
$T\Sigma$, will be denoted by $D_e$. This will be extended to the 
whole ${\mathbb V}^a(\Sigma)$ by requiring $D_et_a=0$. We introduce 
another connection on ${\mathbb V}^a(\Sigma)$, the so-called Sen 
connection \cite{Se} by ${\cal D}_a:=P^b_a\nabla_b$. Clearly, both 
$D_e$ and ${\cal D}_e$ annihilate the fiber metric $g_{ab}$, but the 
projection is annihilated only by $D_e$. (In the language of \cite{KN} 
$D_e$ is a reduction of ${\cal D}_e$, and the reduction is made by 
requiring that the projection be annihilated by the covariant 
derivative operator.) The extrinsic curvature of $\Sigma$ in $M$ is 
$\chi_{ab}:={\cal D}_at_b=\chi_{(ab)}$. In terms of $D_e$ and the 
extrinsic curvature the action of the Sen derivative on an arbitrary 
cross section $X^a$ of ${\mathbb V}^a(\Sigma)$ is given by 

\begin{equation}
{\cal D}_eX^a=D_eX^a+\bigl(\chi_e{}^at_b-t^a\chi_{eb}\bigr)X^b. 
\label{eq:2.2}
\end{equation}
The curvature of ${\cal D}_a$ is defined by the convention $-F^a{}
_{bcd}X^bv^cw^d:=v^c{\cal D}_c(w^d$ ${\cal D}_dX^a)-w^c{\cal D}_c(
v^d{\cal D}_dX^a)-[v,w]^e{\cal D}_eX^a$ for any $X^a$ and $v^c$ and 
$w^c$ tangent to $\Sigma$. This is just the pull back to $\Sigma$ 
of the spacetime curvature 2-form, $F^a{}_{bcd}={}^4R^a{}_{bef}P^e
_cP^f_d$, and it can be re-expressed as 

\begin{eqnarray}
F_{abcd}\!\!\!\!&=\!\!\!\!&R_{abcd}+\chi_{ac}\chi_{bd}-\chi_{ad}\chi
 _{bc}+ \nonumber \\
\!\!\!\!&=\!\!\!\!&t_a\bigl(D_c\chi_{db}-D_d\chi_{cb}\bigr)-
 t_b\bigl(D_c\chi_{da}-D_d\chi_{ca}\bigr), \label{eq:2.3}
\end{eqnarray}
where $R_{abcd}$ is the curvature tensor of the intrinsic geometry 
of $(\Sigma,h_{ab})$. 

${\cal D}_e$ extends in a natural way to the spinor bundle, and its 
action on a spinor field is 

\begin{equation}
{\cal D}_e\lambda_A=D_e\lambda_A-\chi_{eAA'}t^{A'}{}_B\lambda^B. 
 \label{eq:2.4}
\end{equation}
The commutator of two Sen operators acting on the spinor filed 
$\lambda^A$ is

\begin{equation}
\bigl({\cal D}_c{\cal D}_d-{\cal D}_d{\cal D}_c\bigr)\lambda^A=
-F^A{}_{Bcd}\lambda^B-2\chi^e{}_{[c}t_{d]}{\cal D}_e\lambda^A, 
\label{eq:2.5}
\end{equation}
where the curvature $F^A{}_{Bcd}$ is just the pull back to $\Sigma$ 
of the anti-self-dual part of the spacetime curvature 2-form, which 
can also be expressed by the (spinor form of the) intrinsic curvature 
and the extrinsic curvature. For an introduction of the Sen 
connection not using the embedding of $\Sigma$ in $M$, see 
\cite{Sz02}.

The Sen--Witten operator, i.e. the Dirac operator built from the Sen 
connection, is defined to be ${\cal D}:{\mathbb S}^A(\Sigma)
\rightarrow\bar{\mathbb S}_{A'}:\lambda^A\mapsto{\cal D}_{A'A}
\lambda^A$. Since 

\begin{equation*}
\langle{\cal D}_{A'A}\lambda^A,\bar\phi_{B'}\rangle=\int_\Sigma D
_{AA'}\bigl(\lambda^AG^{A'B}\phi_B\bigr){\rm d}\Sigma+\int_\Sigma
\lambda^AG_{AA'}\bigl({\cal D}^{A'B}\phi_B\bigr){\rm d}\Sigma,
\end{equation*}
the formal adjoint of ${\cal D}$ is ${\cal D}^*:\bar{\mathbb S}_{A'}
(\Sigma)\rightarrow{\mathbb S}^A(\Sigma):$ $\bar\phi_{A'}\mapsto
{\cal D}^{AA'}\bar\phi_{A'}$, i.e. essentially the complex conjugate 
of the Sen--Witten operator itself. Therefore, both ${\cal D}^*{\cal 
D}:$ $\lambda^A\mapsto{\cal D}^{AA'}{\cal D}_{A'B}\lambda^B$ and 
${\cal D}{\cal D}^*:$ $\bar\phi_{A'}\mapsto{\cal D}_{A'A}{\cal D}
^{AB'}\bar\phi_{B'}$ are formally self-adjoint and they are 
essentially complex conjugate of each other. Moreover, since 

\begin{eqnarray}
\langle{\cal D}^{AA'}{\cal D}_{A'B}\lambda^B,\phi^C\rangle\!\!\!\!&
 =\!\!\!\!&\int_\Sigma G_{AA'}\bigl({\cal D}^A{}_{B'}\bar\phi^{B'}
\bigr)\bigl({\cal D}^{A'}{}_B\lambda^B\bigr){\rm d}\Sigma+\nonumber \\
\!\!\!\!&+\!\!\!\!&\int_\Sigma D_{AA'}\Bigl(\bigl({\cal D}^{A'}{}_B
 \lambda^B\bigr)G^A{}_{B'}\bar\phi^{B'}\Bigr){\rm d}\Sigma, 
\label{eq:2.6}
\end{eqnarray}
for {\em closed} $\Sigma$ the operator ${\cal D}^*{\cal D}$ is 
positive: $\langle{\cal D}^{AA'}{\cal D}_{A'B}\lambda^B,\lambda^C
\rangle\geq0$ for every spinor field $\lambda^A$.


\subsection{The Sen--Witten identity}
\label{sub-2.3}

Using the commutator (\ref{eq:2.5}), the square of the Sen--Witten 
operator can be written as 

\begin{eqnarray}
{\cal D}_A{}^{A'}{\cal D}_{A'B}\lambda^B\!\!\!\!&=\!\!\!\!&{\cal D}
 _{(A}{}^{A'}{\cal D}_{B)A'}\lambda^B+\frac{1}{2}\varepsilon_{AB}
 {\cal D}_R{}^{R'}{\cal D}_{R'}{}^B\lambda^B= \label{eq:2.7}\\
\!\!\!\!&=\!\!\!\!&-\frac{1}{2}\varepsilon^{A'B'}\bigl({\cal D}
 _{AA'}{\cal D}_{BB'}-{\cal D}_{BB'}{\cal D}_{AA'}\bigr)\lambda^B+
 \frac{1}{2}{\cal D}_e{\cal D}^e\lambda_A=\nonumber \\
\!\!\!\!&=\!\!\!\!&\frac{1}{2}{\cal D}_e{\cal D}^e\lambda_A+
 \frac{1}{2}\varepsilon^{A'B'}F^B{}_{CAA'BB'}\lambda^C+\varepsilon
 ^{A'B'}\chi^e{}_{[a}t_{b]}{\cal D}_e\lambda_A. \nonumber
\end{eqnarray}
The last term can also be written as $\chi^e{}_{AA'}t^{A'}{}_B{\cal 
D}_e\lambda^B$. Using (\ref{eq:2.3}) and the fact that in three 
dimensions the curvature tensor can be expressed by the metric $h
_{ab}$ and the corresponding Ricci tensor and curvature scalar, a 
straightforward computation yields that 

\begin{equation}
\varepsilon^{A'B'}F^B{}_{CAA'BB'}=-\frac{1}{4}\varepsilon_{AC}\bigl(
R+\chi^2-\chi_{de}\chi^{de}\bigr)+\bigl(D_e\chi^e{}_{AA'}-D_{AA'}
\chi\bigr)t^{A'}{}_A. \label{eq:2.8}
\end{equation}
However, the terms on the right hand side are precisely the constraint 
parts of the spacetime Einstein tensor: 

\begin{eqnarray}
\frac{1}{2}\bigl(R+\chi^2-\chi_{ab}\chi^{ab}\bigr)\!\!\!\!&=\!\!\!\!&
 -{}^4G_{ab}t^at^b=\kappa T_{ab}t^at^b=:\kappa \mu, \label{eq:2.9a}\\
\bigl(D_a\chi^a{}_b-D_b\chi\bigr)\!\!\!\!&=\!\!\!\!&-{}^4G_{ae}t^a
 P^e_b=\kappa T_{ae}t^aP^e_b:=\kappa J_b; \label{eq:2.9b} 
\end{eqnarray}
where we used Einstein's field equations. The right hand side of these 
formulae define the energy density and the spatial momentum density of 
the matter fields, respectively, seen by the observer $t^a$. We will 
assume that the matter fields satisfy the dominant energy condition, 
i.e. $\mu^2\geq\vert J_aJ^a\vert$. 
Substituting (\ref{eq:2.8}), (\ref{eq:2.9a}) and (\ref{eq:2.9b}) 
into (\ref{eq:2.7}) finally we obtain 

\begin{eqnarray}
2{\cal D}^{AA'}{\cal D}_{A'B}\lambda^B\!\!\!\!&=\!\!\!\!&{\cal D}_e
 {\cal D}^e\lambda^A+2\chi^{eA}{}_{A'}t^{A'}{}_B{\cal D}_e\lambda^B-
 \nonumber \\
\!\!\!\!&-\!\!\!\!&\frac{1}{2}t_e{}^4G^{ef}t_f\lambda^A+\frac{1}{2}
 t_e\,{}^4G^{ef}P^{AA'}_f2t_{A'B}\lambda^B. \label{eq:2.10}
\end{eqnarray}
This equation is analogous to the Lichnerowicz identity \cite{L}: The 
square of the Dirac operator is expressed in terms of the Laplacian 
and the curvature, but here ${\cal D}_e{\cal D}^e$ is {\em not} the 
intrinsic Laplacian and, in addition, the first derivative of the 
spinor field also appears on the right. Moreover, the curvature in 
(\ref{eq:2.10}) is not simply the scalar curvature, but a genuine 
tensorial piece of that. If, on the other hand, the extrinsic 
curvature is vanishing, then ${\cal D}_e$ reduces to the Levi-Civita 
$D_e$, and (\ref{eq:2.10}) reduces to $2D^{AA'}D_{A'B}\lambda^B=D_e
D^e\lambda^A+\frac{1}{4}R$, which is the genuine Lichnerowicz 
identity for the three dimensional intrinsic Dirac operator. 
It might be interesting to note that the analogous identity for the 
Sen--Witten type operators on two (or more) codimensional submanifolds 
still does {\em not} reduce to the genuine Lichnerowicz identity even 
if the extrinsic curvatures are vanishing, because the reduced 
connection may still have non-trivial curvature in the normal bundle. 
For the example of spacelike 2-surfaces in Lorentzian spacetimes, see 
\cite{Sz07a}. 

Contracting (\ref{eq:2.10}) with $t_{AB'}\bar\phi^{B'}$ and using the 
definitions, equation (\ref{eq:2.4}) and the fact that $G^A{}_{B'}G
^{A'}{}_B$ acts as $-P^a_b$ on vectors tangent to $\Sigma$, we obtain 

\begin{eqnarray}
&{}&D_{AA'}\bigl(2t^A{}_{B'}\bar\phi^{B'}{\cal D}^{A'}{}_B\lambda^B
 \bigr)+2t^{AA'}\bigl({\cal D}_{A'B}\lambda^B\bigr)\bigl({\cal D}
 _{AB'}\bar\phi^{B'}\bigr)= \label{eq:2.11} \\
&=\!\!\!\!&D_a\bigl(\bar\phi^{B'}t_{B'B}{\cal D}^a\lambda^B\bigr)-
 t_{AA'} \bigl({\cal D}_e\lambda^A\bigr)\bigl({\cal D}^e\bar\phi
 ^{A'}\bigr)-\frac{1}{2}t^a\,{}^4G_{aBB'}\lambda^B\bar\phi^{B'}. 
 \nonumber
\end{eqnarray}
Writing the total divergences in a different way we get the Reula--Tod 
(or the $SL(2,{\mathbb C})$ spinor) form \cite{ReTo} of the 
Sen--Witten identity: 

\begin{eqnarray}
D_a\bigl(t^{A'B}\bar\phi^{B'}{\cal D}_{BB'}\lambda^A-\bar\phi^{A'}
 t^{AB'}{\cal D}_{B'B}\lambda^B\bigr)\!\!\!\!&+\!\!\!\!&2t^{AA'}
 \bigl({\cal D}_{A'B}\lambda^B\bigr)\bigl({\cal D}_{AB'}\bar\phi
 ^{B'}\bigr)= \nonumber \\
=-t_{AA'}h^{ef}\bigl({\cal D}_e\lambda^A\bigr)\bigl({\cal D}_f\bar
 \phi^{A'}\bigr)\!\!\!\!&-\!\!\!\!&\frac{1}{2}t^a\,{}^4G_{aBB'}
 \lambda^B\bar\phi^{B'}. \label{eq:2.12}
\end{eqnarray}
Clearly, its right hand side is positive definite for $\lambda^A=
\phi^A$ and matter fields satisfying the dominant energy condition. 
This identity is the basis of (probably the simplest) proof of the 
positivity of the ADM and Bondi--Sachs energies. (For the original 
proofs using Dirac spinors, see \cite{Wi,Ne}, and for its extension 
to include black holes, see \cite{GHHP,ReTo}.) The basic idea is that 
if $\Sigma$ is asymptotically flat and $\lambda^A=\phi^A$ is chosen 
to be an asymptotically constant solution to the Sen-Witten equation 
${\cal D}_{A'A}\lambda^A=0$, then the second term on the left is 
vanishing, and then, taking the integral of (\ref{eq:2.12}) and 
converting the total divergence to a 2-surface integral at infinity, 
the left hand side gives the ${}_0\lambda^A{}_0\bar\lambda
^{A'}$-component of the ADM energy-momentum, where ${}_0\lambda^A$ is 
the asymptotic value of the spinor field $\lambda^A$. (At null 
infinity $\lambda^A$ cannot be required to be asymptotically constant, 
only a weaker boundary condition may be imposed. For the details see 
\cite{ReTo}.)


\section{The eigenvalue problem for the Sen--Witten operators}
\label{sec-3}

According to the general theory of spinors (see e.g. the appendix of 
\cite{PRII}) in three dimensions the spinors have two components, 
moreover the Sen--Witten operator maps cross sections of ${\mathbb 
S}^A(\Sigma)$ to cross sections of the complex conjugate bundle $\bar
{\mathbb S}_{A'}(\Sigma)$, it seems natural to define the eigenvalue 
problem by 

\begin{equation}
{\rm i}G_A{}^{A'}{\cal D}_{A'}{}^B\psi_B=-\frac{1}{\sqrt{2}}\beta
\psi_A. \label{eq:3.1}
\end{equation}
The unitary spinor form \cite{Re,Fr} of (\ref{eq:3.1}), namely ${\rm i}
{\cal D}_A{}^B\psi_B=-\frac{1}{\sqrt2}\beta\psi_A$, apparently makes 
this definition of the eigenvalue problem reasonable. (The choice 
for the apparently ad hoc coefficient $-1/\sqrt{2}$ in front of the 
eigenvalue $\beta$ yields the compatibility with the known standard 
results in special cases.) 
However, it is desirable that the Hermitian metric be compatible with 
the connection in the sense that ${\cal D}_eG_{AA'}=0$. Unfortunately, 
since ${\cal D}_eG_{AA'}$ is $\sqrt2$-times the extrinsic curvature of 
$\Sigma$, in general this requirement cannot be satisfied. As a 
consequence, in general the eigenvalue $\beta$ is {\em not} real. 
In fact, a straightforward calculation (by elementary integration by 
parts) gives that 

\begin{equation}
\beta\Vert\psi_A\Vert^2=\bar\beta\Vert\psi_A\Vert^2+{\rm i}\int
_\Sigma\chi G_{AA'}\psi^A\bar\psi^{A'}{\rm d}\Sigma+{\rm i}\sqrt{2}
\int_\Sigma D_{AA'}\bigl(\psi^A\bar\psi^{A'}\bigr){\rm d}\Sigma. 
\label{eq:3.2}
\end{equation}
This implies that, even if $\Sigma$ is closed, which will be assumed 
in the rest of this paper, the imaginary part of $\beta$ is 
proportional to the integral of mean curvature $\chi$ weighted by the 
pointwise norm $G_{AA'}\psi^A\bar\psi^{A'}$, which is not zero in 
general. 

This difficulty raises the question whether we can find a slightly 
different definition of the eigenvalue problem for the Sen--Witten 
operator yielding {\em real} eigenvalues. To motivate this, observe 
that although the base manifold $\Sigma$ is only three dimensional, 
the connection ${\cal D}_e$ is four dimensional in its spirit, as 
originally it is defined on the Lorentzian vector bundle ${\mathbb 
V}^a(\Sigma)$. Since its fibres are four dimensional, the 
corresponding spinors are the four component Dirac spinors. Hence 
we should define the eigenvalue problem for the Sen--Witten operator 
in terms of the Dirac spinors. 

Recall that a Dirac spinor $\Psi^\alpha$ is a pair of Weyl spinors 
$\lambda^A$ and $\bar\mu^{A'}$, written them as a column vector 

\begin{equation}
\Psi^\alpha=\left(\begin{array}{cc}\lambda^A \\  
                          \bar\mu^{A'}\end{array}\right) 
\label{eq:3.3}
\end{equation}
and adopting the convention $\alpha=A\oplus{A'}$, $\beta=B\oplus{B'}$ 
etc. Its derivative ${\cal D}_e\Psi^\alpha$ is the column vector 
consisting of ${\cal D}_e\lambda^A$ and ${\cal D}_e\bar\mu^{A'}$. If 
Dirac's $\gamma$-`matrices' are denoted by $\gamma^\alpha_{e\beta}$, 
then one can consider the eigenvalue problem 

\begin{equation}
{\rm i}\gamma^\alpha_{e\beta}{\cal D}^e\Psi^\beta=\alpha\Psi^\alpha.  
\label{eq:3.4}
\end{equation}
Explicitly, with the representation 

\begin{equation}
\gamma^\alpha_{e\beta}=\sqrt{2}\left(\begin{array}{cc}
          0&\varepsilon_{E'B'}\delta^A_E \\
      \varepsilon_{EB}\delta^{A'}_{E'}&0 \end{array}\right) 
\label{eq:3.5}
\end{equation}
(see e.g. \cite{PRI}, pp 221), this is just the pair of equations 

\begin{equation}
{\rm i}{\cal D}_{A'}{}^A\lambda_A=-\frac{\alpha}{\sqrt2}\bar\mu_{A'}, 
\hskip 20pt
{\rm i}{\cal D}_A{}^{A'}\bar\mu_{A'}=-\frac{\alpha}{\sqrt2}\lambda_A.
\label{eq:3.6}
\end{equation}
These imply that both the unprimed and the primed Weyl spinor parts 
of $\Psi^\alpha$ are eigenspinors of the square of the Sen--Witten 
operator with the {\em same} eigenvalue: 

\begin{equation}
2{\cal D}^{AA'}{\cal D}_{A'B}\lambda^B=\alpha^2\lambda^A, 
\hskip 20pt
2{\cal D}^{A'A}{\cal D}_{AB'}\bar\mu^{B'}=\alpha^2\bar\mu^{A'}.
\label{eq:3.7}
\end{equation}
Then by (\ref{eq:2.6}) $0\leq2\langle{\cal D}^{AA'}{\cal D}_{A'B}
\lambda^B,\lambda^C\rangle=\alpha^2\Vert\lambda^A\Vert$, i.e. {\em 
the eigenvalues $\alpha$ are real}. Conversely, if the pair $(\alpha
^2,\lambda^A)$ is a solution of the eigenvalue problem for $2{\cal D}
^*{\cal D}$ with nonzero $\alpha$, then $(\pm\alpha,\Psi^\alpha_\pm)$ 
with $\bar\mu^{A'}:=\mp(\sqrt{2}/\alpha){\rm i}{\cal D}^{A'A}\lambda
_A$ are solutions of the eigenvalue problem (\ref{eq:3.4}). Therefore, 
it is enough to study the eigenvalue problem for the second order 
operator $2{\cal D}^*{\cal D}$. 

By (\ref{eq:3.3}) $\Psi^\alpha=(\lambda^A,\bar\mu^{A'})$ is a Dirac 
eigenspinor with eigenvalue $\alpha$ precisely when $(\lambda^A,-
\bar\mu^{A'})$ is a Dirac eigenspinor with eigenvalue $-\alpha$. In 
the language of Dirac spinors this is formulated in terms of the 
chirality, represented by the so-called `$\gamma_5$-matrix', denoted 
here by 

\begin{equation}
\eta^\alpha{}_\beta:=\frac{1}{4!}\varepsilon^{abcd}\gamma^\alpha
_{a\mu}\gamma^\mu_{b\nu}\gamma^\nu_{c\rho}\gamma^\rho_{d\beta}=
{\rm i}\left(\begin{array}{cc} \delta^A_B&0\\ 
        0&-\delta^{A'}_{B'}\end{array}\right) 
\label{eq:3.8}
\end{equation}
(see appendix II. of \cite{PRII}). Since this is anti-commuting with 
$\gamma^\alpha_{e\beta}$, from (\ref{eq:3.4}) we obtain that ${\rm i}
\gamma^\alpha_{e\mu}{\cal D}^e(\eta^\mu{}_\beta\Psi^\beta)=-\alpha
(\eta^\alpha{}_\beta\Psi^\beta)$. Thus if $\Psi^\alpha$ is a Dirac 
eigenspinor with eigenvalue $\alpha$, then, in fact, $\eta^\alpha{}
_\beta\Psi^\beta$ is a Dirac eigenspinor with eigenvalue $-\alpha$. 

On the other hand, if there are Dirac eigenspinors with definite 
chirality, then they  belong to the kernel of the Sen--Witten 
operator. Indeed, Dirac spinors with definite chirality have the 
structure either $(\lambda^A,0)$ or $(0,\bar\mu^{A'})$, which, by 
(\ref{eq:3.6}), yield that ${\cal D}_{A'A}\lambda^A=0$ or ${\cal D}
_{AA'}\bar\mu^{A'}=0$, respectively. Therefore, this notion of 
chirality cannot be used to decompose the space of the eigenspinors 
with given eigenvalue. Its role is simply to take a Dirac eigenspinor 
with eigenvalue $\alpha$ to a Dirac eigenspinor with eigenvalue 
$-\alpha$. 

By the reality of the eigenvalues both the complex conjugate of the 
unprimed spinor part $\lambda^A$ and the primed spinor part $\bar
\mu^{A'}$ of $\Psi^\alpha$ are eigenspinors of $2{\cal D}{\cal D}^*$ 
with the same eigenvalue $\alpha^2$. This raises the question as 
whether the eigenvalue problem can be restricted by $\lambda^A=\mu
^A$, i.e. by requiring the Dirac eigenspinors $\Psi^\alpha$ to be 
Majorana spinors. However, (\ref{eq:3.6}) implies that in this case 
$\alpha$ would have to be purely imaginary or zero, i.e. the 
Sen--Witten operator does not have genuine, non-trivial Majorana 
eigenspinors. 

Finally suppose that the extrinsic curvature is vanishing. In this 
special case ${\cal D}_e=D_e$, and let us consider the eigenvalue 
problem defined by (\ref{eq:3.1}). Then ${\rm i}G_A{}^{A'}D_{A'}{}
^B({\rm i}G_B{}^{B'}D_{B'}{}^C\psi_C)=\frac{1}{2}\beta^2\psi_A$. 
However, by $D_eG_{AA'}=0$ we can write 

\begin{eqnarray*}
\beta^2\psi_A\!\!\!\!&=\!\!\!\!&-2G_A{}^{A'}G^B{}_{B'}D_{A'B}\bigl(
 D^{B'C}\psi_C\bigr)=-2G_A{}^{A'}G_{A'}{}^BD_{BB'}\bigl(D^{B'C}\psi
 _C\bigr)= \\
\!\!\!\!&=\!\!\!\!&-2G_{AA'}G^{A'B}\bigl(D_B{}^{B'}D_{B'}{}^C\psi_C
 \bigr)=-2D_A{}^{A'}D_{A'}{}^B\psi_B.
\end{eqnarray*}
Thus the pair $(\beta,\psi^A)$ is a solution of the eigenvalue 
problem for $D^*D$, and hence we may write $\beta=\alpha$ and $\psi
^A=\lambda^A$. Then $\alpha\bar\mu_{A'}=-{\rm i}\sqrt{2}D_{A'}{}^A
\lambda_A={\rm i}\sqrt{2}G_{A'}{}^AG_A{}^{B'}D_{B'}{}^B\lambda_B=
{\rm i}\sqrt{2}G_{A'}{}^A(\frac{1}{\sqrt2}{\rm i}\alpha\lambda_A)
=\alpha G_{A'A}\lambda^A$; i.e. the primed spinor part $\bar\mu
_{A'}$ of the Dirac eigenspinor is just $G_{A'A}\lambda^A$. 
Therefore, in the special case of the vanishing extrinsic curvature 
the eigenvalue problems (\ref{eq:3.1}) and (\ref{eq:3.4}) coincide.


\section{Lower bounds for the eigenvalues}
\label{sec-4}

Suppose that $\lambda^A$ is an eigenspinor of $2{\cal D}^*{\cal D}$ 
with eigenvalue $\alpha^2$. Then since we assumed that $\Sigma$ is 
closed, (\ref{eq:2.12}) yields that 

\begin{eqnarray}
\alpha^2\Vert\lambda^A\Vert^2\!\!\!\!&=\!\!\!\!&2\sqrt{2}\int_\Sigma
 \bigl(t_{A'A}\bar\lambda^{A'}{\cal D}^{AB'}{\cal D}_{B'B}\lambda^B
 \bigr){\rm d}\Sigma=\label{eq:4.1} \\
\!\!\!\!&=\!\!\!\!&\sqrt{2}\int_\Sigma\Bigl(-t_{AA'}\bigl({\cal D}_e
 \lambda^A\bigr)\bigl({\cal D}^e\bar\lambda^{A'}\bigr)-\frac{1}{2}
 t^a\,{}^4G_{aBB'}\lambda^B\bar\lambda^{B'}\Bigr){\rm d}\Sigma.  
 \nonumber 
\end{eqnarray}
This gives a lower bound for the eigenvalue $\alpha^2$: 

\begin{equation}
\alpha^2\geq-\frac{1}{\sqrt{2}\Vert\lambda^A\Vert^2} \int_\Sigma t^a
\,{}^4G_{aBB'}\lambda^B\bar\lambda^{B'}\,{\rm d}\Sigma\geq-\frac{1}{2}
\inf\frac{\int_\Sigma t^a\,{}^4G_{ab}l^b\,{\rm d}\Sigma}{\int_\Sigma 
t_bl^b\,{\rm d}\Sigma}, \nonumber
\end{equation}
where the infimum is taken on the set of the smooth, future pointing 
null vector fields $l^a$ on $\Sigma$. However, this bound is certainly 
{\em not} sharp: In the special case of the vanishing extrinsic 
curvature the nominator is the integral of $-\frac{1}{2}Rt_al^a$ (see 
equations (\ref{eq:2.9a})-(\ref{eq:2.9b})), yielding Lichnerowicz's 
bound $\frac{1}{4}\inf\{ R(p)\vert p\in\Sigma\}$ instead of 
Friedrich's sharp bound $\frac{3}{8}\inf\{ R(p)\vert p\in\Sigma\}$. 

To find the sharp bound, we follow the general philosophy of 
\cite{TFr,TF00} (see also \cite{Zh}) and consider the modified Sen 
connection 

\begin{equation}
\tilde{\cal D}_e\lambda^A:={\cal D}_e\lambda^A+sP^{AA'}_e{\cal D}
_{A'B}\lambda^B \label{eq:4.2}
\end{equation}
for some real constant $s$. Then a straightforward calculation gives 

\begin{eqnarray}
t_{AA'}\bigl(\tilde{\cal D}_e\lambda^A\bigr)\bigl(\tilde{\cal D}^e
 \bar\lambda^{A'}\bigr)\!\!\!\!&+\!\!\!\!&2s(1+\frac{s}{4})t^{AA'}
 \bigl({\cal D}_{AB'}\bar\lambda^{B'}\bigr)\bigl({\cal D}_{A'B}
 \lambda^B\bigr)= \nonumber \\
\!\!\!\!&=\!\!\!\!&t_{AA'}\bigl({\cal D}_e\lambda^A\bigr)\bigl({\cal 
 D}^e\bar\lambda^{A'}\bigr). \nonumber 
\end{eqnarray}
Using this expression for $t_{AA'}({\cal D}_e\lambda^A)({\cal D}^e
\bar\lambda^{A'})$ in (\ref{eq:4.1}) we obtain 

\begin{equation}
(1+s+\frac{3}{4}s^2)\alpha^2\Vert\lambda^A\Vert^2=\sqrt{2}\int
_\Sigma\Bigl(-t_{AA'}\bigl(\tilde{\cal D}_e\lambda^A\bigr)\bigl(
\tilde{\cal D}^e\bar\lambda^{A'}\bigr)-\frac{1}{2}t^a\,{}^4G_{aBB'}
\lambda^B\bar\lambda^{B'}\Bigr){\rm d}\Sigma. \label{eq:4.3} 
\end{equation}
Its left hand has a minimum at $s=-\frac{2}{3}$, in which case 

\begin{equation}
\alpha^2\Vert\lambda^A\Vert^2=\frac{3}{\sqrt{2}}\int_\Sigma\Bigl(-t
_{AA'}\bigl(\tilde{\cal D}_e\lambda^A\bigr)\bigl(\tilde{\cal D}^e
\bar\lambda^{A'}\bigr)-\frac{1}{2}t^a\,{}^4G_{aBB'}\lambda^B\bar
\lambda^{B'}\Bigr){\rm d}\Sigma. \label{eq:4.4} 
\end{equation}
Remarkably enough, apart from the numerical coefficient $\frac{3}
{\sqrt2}$ the right hand side is precisely the integral of the right 
hand side of (\ref{eq:2.12}), whose integral on an asymptotically 
flat $\Sigma$ gave the appropriate component of the total 
energy-momentum of the localized matter+gravity systems. In fact, 
using the unitary spinor form ${\cal D}_{EF}:=G_F{}^{E'}{\cal D}
_{E'E}={\cal D}_{(EF)}$ of the Sen derivative operator ${\cal D}_e$ 
the decomposition of the derivative ${\cal D}_e\lambda_A$ into its 
irreducible parts is 

\begin{eqnarray}
G_F{}^{E'}{\cal D}_{E'E}\lambda_A\!\!\!\!&=\!\!\!\!&{\cal D}_{(EF}
 \lambda_{A)}+\frac{1}{3}\varepsilon_{EA}{\cal D}_{FB}\lambda^B+
 \frac{1}{3}\varepsilon_{FA}{\cal D}_{EB}\lambda^B= \label{eq:4.5} \\
\!\!\!\!&=\!\!\!\!&{\cal D}_{(EF}\lambda_{A)}+\frac{1}{3}\varepsilon
 _{EA}G_F{}^{K'}{\cal D}_{K'K}\lambda^K+\frac{1}{3}\varepsilon_{FA}
 G_E{}^{K'}{\cal D}_{K'K}\lambda^K= \nonumber \\
\!\!\!\!&=\!\!\!\!&{\cal D}_{(EF}\lambda_{A)}+\frac{1}{3}G_F{}^{E'}
 \bigl(\varepsilon_{EA}\delta^{K'}_{E'}-G_E{}^{K'}G_{E'A}\bigr){\cal 
 D}_{K'K}\lambda^K= \nonumber \\
\!\!\!\!&=\!\!\!\!&{\cal D}_{(EF}\lambda_{A)}+\frac{2}{3}G_F{}^{E'}
 P^{KK'}_{EE'}\varepsilon_{KA}{\cal D}_{K'L}\lambda^L; \nonumber
\end{eqnarray}
and hence, taking into account that in the positive energy proofs 
${\cal D}_{A'A}\lambda^a=0$, 

\begin{eqnarray}
-t_{AA'}\bigl(\tilde{\cal D}_e\lambda^A\bigr)\bigl(\tilde{\cal D}^e
 \bar\lambda^{A'}\bigr)\!\!\!\!&=\!\!\!\!&2t^{AA'}t^{BB'}t^{EE'}
 \bigl({\cal D}_{(AB}\lambda_{E)}\bigr)\bigl({\cal D}_{(A'B'}\bar
 \lambda_{E')}\bigr)= \label{eq:4.6} \\
\!\!\!\!&=\!\!\!\!&-t_{AA'}\bigl({\cal D}_e\lambda^A\bigr)\bigl(
 {\cal D}^e\bar\lambda^{A'}\bigr). \nonumber
\end{eqnarray}
It might be interesting to note that ${\cal D}_{(AB}\lambda _{C)}$ 
is just the 3-surface twistor derivative of the spinor field 
\cite{Tod,PRII,Sz00}: ${\cal D}_{(AB}\lambda _{C)}=0$ is the purely 
spatial part in the complete irreducible 3+1 decomposition of the 
1-valence spacetime twistor equation $\nabla_{A'(A}\lambda_{B)}=0$. 

If we introduced the derivative $\tilde{\cal D}_e\lambda^A$ by the 
more general expression ${\cal D}_e\lambda^A+sP^{AA'}_e{\cal D}_{A'B}
\lambda^B+FP^{AA'}_et_{A'B}\lambda^B$ for some real constant $s$ and 
complex function $F$, then in the integrand on the right hand side 
of (\ref{eq:4.4}) we would have the extra {\em negative definite} 
term $-\frac{3}{4}F\bar Ft_{AA'}\lambda^A\bar\lambda^{A'}$. This 
term would {\em decrease} the right hand side of (\ref{eq:4.4}) (and 
the lower bound below), and hence its introduction does not seem to 
be useful. 

Finally, by (\ref{eq:4.4}) we have the lower bound for the eigenvalues 

\begin{equation}
\alpha^2\geq-\frac{3}{4}\inf\frac{\int_\Sigma t^a\,{}^4G_{ab}l^b\,
{\rm d}\Sigma}{\int_\Sigma t_bl^b\,{\rm d}\Sigma}=\frac{3}{4}\kappa
\inf\frac{\int_\Sigma t^aT_{ab}l^b\,{\rm d}\Sigma}{\int_\Sigma t_b
l^b\,{\rm d}\Sigma}, \label{eq:4.7}
\end{equation}
where, as above, the infimum is taken on the set of the smooth, 
future pointing null vector fields $l^a$ on $\Sigma$. The quotient of 
the integrals is some average on $\Sigma$ of the flux of the energy 
current $T^a{}_bl^b$ of the matter fields seen by the null observer 
$l^a$. In the special case of the vanishing extrinsic curvature this 
bound is not less then Friedrich's sharp lower bound.


\section{The limiting case}
\label{sec-5}

If the equality hols in (\ref{eq:4.7}), then by (\ref{eq:4.4}) and 
(\ref{eq:4.6}) the eigenspinor $\lambda_A$ must also solve the 
3-surface twistor equation ${\cal D}_{(AB}\lambda_{C)}=0$. Then the 
derivative of the spinor field $\lambda_A$ can be expressed in terms 
of $\bar\mu_{A'}$ algebraically, and we can evaluate its 
integrability condition to obtain a condition on the geometry of the 
data set $(\Sigma,h_{ab},\chi_{ab})$. However, instead of the general 
analysis of this limiting case we show directly through an example 
that the lower bound (\ref{eq:4.7}) is sharp. 

The example is the $t={\rm const}$ spacelike hypersurface in a $k=1$ 
Friedman--Robertson--Walker cosmological spacetime. Explicitly, the 
manifold $\Sigma$ is homeomorphic to $S^3$, the intrinsic metric $h
_{ab}$ is the standard 3-sphere metric with scalar curvature $R={\rm 
const}$, and the extrinsic curvature is $\chi_{ab}=\frac{1}{3}\chi 
h_{ab}$ with $\chi={\rm const}$. For this data set $t^a\,{}^4G_{ab}
P^b_c=0$ and $-t^at^b\,{}^4G_{ab}=\frac{1}{2}R+\frac{1}{3}\chi^2={\rm 
const}$, and hence the lower bound (\ref{eq:4.7}) is $\frac{3}{8}R+
\frac{1}{4}\chi^2$. 

On the other hand, we know that this example with $\chi=0$ saturates 
the inequality of Friedrich, i.e. the smallest eigenvalue ${}_0\alpha
^2_1$ of the (Riemannian) eigenvalue problem $2D^{AA'}D_{A'B}\lambda
^B={}_0\alpha^2\lambda^A$ is just $\frac{3}{8}R$. We show that the 
corresponding eigenspinor is an eigenspinor of $2{\cal D}^{AA'}{\cal D}
_{A'B}$ too, and the corresponding eigenvalue saturates (\ref{eq:4.7}). 
In fact, since $\chi={\rm const}$, $2{\cal D}^{AA'}{\cal D}_{A'B}
\lambda^B=2D^{AA'}D_{A'B}\lambda^B+\frac{1}{4}\chi^2\lambda^A$ holds, 
and hence for the smallest eigenvalue $\alpha_1$ of the Sen--Witten 
operator we obtain $\alpha^2_1={}_0\alpha^2_1+\frac{1}{4}\chi^2$, just 
the lower bound coming from (\ref{eq:4.7}). The extrinsic curvature 
shifted both Friedrich's lower bound and the smallest Riemannian 
eigenvalue by the same positive term $\frac{1}{4}\chi^2$. It is easy 
to see that the 3-surface twistor operator also annihilates this 
eigenspinor: since it is annihilated by the Riemannian 3-surface 
twistor operator and ${\cal D}_{AB}\lambda_C=D_{AB}\lambda_C+
\frac{1}{6\sqrt{2}}\chi(2\varepsilon_{BC}\lambda_A+\varepsilon_{AB}
\lambda_C)$ holds, ${\cal D}_{(AB}\lambda_{C)}=0$ follows.


\hskip 25pt

This work was partially supported by the Hungarian Scientific Research 
Fund (OTKA) grant K67790.


\noindent

\begin{thebibliography}{9999}

\bibitem{L} A. Lichnerowicz, Spineurs harmoniques, C. R. Acad. Sci. 
          Paris A--B {\bf 257} 7--9 (1963)

\bibitem{TFr} Th. Friedrich, Der erste Eigenwert des Dirac-operators 
         einer kompakten Riemannschen Mannigfaltigkeit nichtnegativer 
         Skalarkr\"ummung, Math. Nachr. {\bf 97} 117--146 (1980)

\bibitem{Hi86} O. Hijazi, A conformal lower bound for the smallest 
         eigenvalue of the Dirac operator and Killing spinors, Commun. 
         Math. Phys. {\bf 104} 151--162 (1986)

\bibitem{Hi95} O. Hijazi, Lower bounds for the eigenvalues of the 
         Dirac operator, J. Geom. Phys. {\bf 16} 27--38 (1995) 

\bibitem{TF00} Th. Friedrich, {\it Dirac Operators in Riemannian 
         Geometry}, Graduate Studies in Mathematics, Vol 25, AMS 
         Providence, Rhode Island 2000

\bibitem{Ba92} C. B\"ar, Lower eigenvalue estimates for Dirac 
         operators, Math. Ann. {\bf 293} 39--46 (1992) 

\bibitem{FK} Th. Friedrich, E. C. Kim, Some remarks on the Hijazi 
         inequality and generalizations of the Killing equation for 
         spinors, J. Geom. Phys. {\bf 37} 1--14 (2001)

\bibitem{Zh} X. Zhang, Lower bounds for eigenvalues of hypersurface 
          Dirac operators, Math. Res. Lett. {\bf 5} 199--210 (1998)

\bibitem{PRI}
         R. Penrose, W. Rindler, {\it Spinors and Spacetime}, Vol 1, 
          Cambridge University Press, Cambridge 1984


\bibitem{Se} A. Sen, On the existence of neutrino `zero-modes' in 
         vacuum spacetimes, J. Math. Phys. {\bf 22} 1781--1786 (1981) 

\bibitem{KN} S. Kobayashi, K. Nomizu, {\it Foundations of Differential 
         Geometry}, vol. 1, Interscience, New York, 1964 

\bibitem{Sz02} L. B. Szabados, On the role of conformal three-geometries 
         in the dynamics of general relativity, Class. Quantum Grav. 
         {\bf 19} 2375--2391 (2002), gr-qc/0110106


\bibitem{Sz07a} L. B. Szabados, Total angular momentum from Dirac 
         eigenspinors, Class. Quantum Grav. (to appear), arXiv:0709.1072


\bibitem{ReTo} O. Reula, K. P. Tod, Positivity of the Bondi energy, 
         J. Math. Phys. {\bf 25} 1004--1008 (1984) 

\bibitem{Wi} E. Witten, A new proof of the positive energy theorem, 
          Commun. Math. Phys. {\bf 30} 381--402 (1981)

\bibitem{Ne} J. M. Nester, A new gravitational energy expression and 
          with a simple positivity proof, Phys. Lett. A, {\bf 83} 
          241--242 (1981)

\bibitem{GHHP} G. W. Gibbons, S. W. Hawking, G. T. Horowitz, M. J. 
          Perry, Positive mass theorem for black holes, Commun. Math. 
          Phys. {\bf 88} 295--308 (1983)

\bibitem{Re} O. Reula, Existence theorem for solutions of Witten's 
          equation and nonnegativity of total mass, J. Math. Phys. 
          {\bf 23} 810--814 (1982)

\bibitem{Fr} J. Frauendiener, Triads and the Witten equation, Class. 
           Quantum Gravity, {\bf 8} 1881--1187 (1991) 

\bibitem{Tod} K. P. Tod, Three-surface twistors and conformal 
         embedding, Gen. Rel. Grav. {\bf 16} 435--443 (1984) 

\bibitem{PRII}
         R. Penrose, W. Rindler, {\it Spinors and Spacetime}, Vol 2, 
          Cambridge University Press, Cambridge 1986

\bibitem{Sz00} L. B. Szabados, On certain global conformal invariant 
         and 3-surface twistors of initial data sets, Class. Quantum 
         Grav. {\bf 17} 793--811 (2000), gr-qc/9909052





\end{thebibliography}
\end{document}